\documentclass[a4paper,
twocolumn,
amsmath,amssymb,superscriptaddress,floatfix,nofootOAinbib]{revtex4-1}
\usepackage{graphicx}
\usepackage{dcolumn}
\usepackage{bm}
\usepackage{xcolor}
\usepackage{url}
\usepackage{longtable}

\usepackage{subfigure}
\usepackage{multirow}

\usepackage{booktabs}
\usepackage{tabularx}

\begin{document}
\title{Maximum Entropy Method for Valence Quark Distributions in Exotic Hadrons: A Study of the Zc(3900) Case}

\author{Chengdong Han}
\email{chdhan@impcas.ac.cn}
\affiliation{Institute of Modern Physics,Chinese Academy of Sciences, Lanzhou 730000, China}
\affiliation{School of Nuclear Science and Technology, University of Chinese Academy of Sciences, Beijing 100049, China}

\author{Xiaopeng Wang}
\affiliation{Institute of Modern Physics,Chinese Academy of Sciences, Lanzhou 730000, China}
\affiliation{School of Nuclear Science and Technology, University of Chinese Academy of Sciences, Beijing 100049, China}
\author{Wei Kou}
\affiliation{Institute of Modern Physics,Chinese Academy of Sciences, Lanzhou 730000, China}
\affiliation{School of Nuclear Science and Technology, University of Chinese Academy of Sciences, Beijing 100049, China}

\collaboration{These authors contributed equally to this work.}

\author{Xurong Chen}
\email{xchen@impcas.ac.cn (corresponding\,author)}
\affiliation{Institute of Modern Physics,Chinese Academy of Sciences, Lanzhou 730000, China}
\affiliation{School of Nuclear Science and Technology, University of Chinese Academy of Sciences, Beijing 100049, China}
\begin{abstract}
 In this study we demonstrate the application of the Maximum Entropy Method (MEM) to determine the valence quark distribution of exotic hadrons. Our investigation yields three key findings. Firstly, we observe a significant shift towards smaller Bjorken scale $x$ in the peak position of the valence quark distribution for hadrons with an increasing number of valence quarks, consistent with previous results by Kawamura and Kumano. Secondly, assuming that the $Z_c(3900)$ initially consists of four valence quarks, we employ MEM to determine its initial valence quark distribution, estimating a radius of $r_c=1.276$ fm at an extremely low resolution scale $Q^2$. Furthermore, we identify a notable discrepancy between our computed charge form factor $G_c(q)$ at leading order and the outcomes of hadron molecular state calculations. We propose that this form factor can be extracted from the QCD counting rule cross-section, which is grounded in Generalized Distribution Amplitudes (GDA) linked to the multi-quark states.

\end{abstract}


\maketitle

\section{Introduction}
\label{SecI}
As we know, the quark model classifies hadrons into two categories: mesons with a quark and an antiquark, and baryons with three quarks. Within the framework of not violating the fundamental theory of strong interactions, quantum chromodynamics (QCD), there are also speculations about other hadron configurations, such as tetraquark hadrons with a $qq\bar{q}\bar{q}$ configuration and pentaquark hadrons with a $qqqq\bar{q}$ configuration. Tetraquark and pentaquark particles are thought to have been observed and are under investigation. However, hexaquarks consisting of either a dibaryon or three quark-antiquark pairs have not yet been confirmed as observed. This is called a strange hadron, which is a subatomic particle composed of quarks and gluons, but unlike the traditional well-known hadrons such as the proton, neutron, and meson, it is composed of more than three valence quarks. The exotic hadron belongs to a type of matter whose existence is not predicted by the standard model of particle physics.

In recent years, due to the discovery of more and more exotic hadrons, such as the $X$(3872) discovered in 2003 by the Belle experiment \cite{Belle:2003nnu} and later confirmed by several other experimental collaborations, the $Z^{+}(4430)$ \cite{Belle:2007hrb} discovered in the Belle experiment and confirmed by the LHCb Collaboration, and the $P^{+}_{c}(4380)$ and $P^{-}_{c}(4450)$ \cite{LHCb:2015yax} discovered by the LHCb Collaboration, studying the nature of exotic hadron particles has become an extremely important problem in hadron physics. The $Z_{c}(3900)$ is believed to be the first tetraquark observed experimentally. The tetraquark states $Z_{c}(3900)$ were discovered by two independent research groups (BESIII and Belle Collaborations) in 2013 \cite{BESIII:2013ris, Belle:2013yex, Xiao:2013iha, Wilbring:2013cha}. There is still a lot of debate about the possible internal structure, with some interpretations suggesting a molecular $\bar{D}^{*}D$ state \cite{Wilbring:2013cha, Wang:2013cya, Guo:2013sya, Dong:2013iqa, Zhang:2013aoa, Ke:2013gia, Aceti:2014uea, He:2015mja}, or a tetraquark of various configurations \cite{Braaten:2013boa, Dias:2013xfa, Maiani:2014aja, Esposito:2014rxa, Qiao:2013raa, Wang:2013vex, Deng:2014gqa}. The internal structure of exotic hadrons has recently become a hot area for investigating the longitudinal partonic structure, with studies focusing on the parton longitudinal momentum distributions of hadrons through various processes, including deep inelastic scattering of charged leptons and neutrinos, hadron productions, $W$ and $Z$ boson jet, and Drell-Yan processes. Therefore, the internal structure of exotic hadrons has become a hot area for investigating the longitudinal partonic structure. Recently, Kawamura and Kumano \cite{Kawamura:2013wfa} proposed that the internal structure of exotic hadron candidates can be determined by high-energy reactions with exclusive processes. In this study, we consider the $Z_{c}(3900)$ to be a tetraquark state to obtain the valence quark distributions using the maximum entropy method (MEM).

Information entropy theory provides a constructive criterion for establishing maximum probability distributions under known constraints \cite{Jaynes:1957zza}. The concept of maximum entropy originated in the statistical physics of Boltzmann and Gibbs as a theoretical tool for predicting the equilibrium states of thermal systems. Maximum Entropy serves as a unique and reliable method for ensuring basic consistency axioms when making inferences about distribution functions based on observations. Boltzmann and Gibbs utilized it for predicting material equilibria, Shannon for computing channel capacities for information transmission, and Jaynes for framing physical problems as matters of inference.
The MEM provides the least biased estimate possible given the available information and is extensively utilized in Lattice QCD (LQCD) \cite{Asakawa:2000tr, Ding:2011hr}, offering reliable theoretical results and high computational efficiency.
In summary, MEM has a broad range of applications across various research fields \cite{presse2013principles, Jaynes:1957zza, Jaynes:1957zz, martyushev2006maximum}. Particularly in recent times, it has been used to determine the initial valence quark distribution of hadrons such as the proton \cite{Wang:2014lua, Han:2016bsw}, neutron \cite{Han:2018bxt}, pion \cite{Han:2018wsw}, and kaon \cite{Han:2020vjp}.

In this study, we propose using the MEM to determine the valence quark distributions of exotic hadrons and to summarize the rules governing the valence quark momentum distribution in hadrons with varying numbers of quarks. Additionally, we aim to calculate the valence quark distribution function and radius of the $Z_{c}(3900)$ for the first time using MEM, based on known structural information and properties of the $Z_{c}(3900)$ in the naive quark model and QCD theory. In this analysis, we consider only the electromagnetic interaction among the four quarks in the $Z_{c}(3900)$ to define the spatial region for each valence quark to occupy. Subsequently, we can derive the structure function and the leading-order charge form factor $G^{LO}_{c}(q)$ of the $Z_{c}(3900)$. We anticipate that this form factor can be extracted from the QCD counting rule cross-section, constructed using the generalized distribution amplitude (GDA) associated with the multi-quark state.

\section{Determination of exotic hadron valance quark distributions from maximum entropy method}
\label{SecII}
The simplest representation of the internal structure of exotic hadrons is provided by the quark model, where the exotic hadron is composed of multiple valence quarks. Therefore, the parametrization of the nonperturbative input \cite{Pumplin:2002vw} for exotic hadrons can be expressed as follows:
\begin{equation}
\begin{aligned}
f_v(x,Q_0^2)=A_f x^{B_f}(1-x)^{C_f}.
\end{aligned}
\label{Parametrization_hadron}
\end{equation}
In the quark model, an exotic hadron consists of $n_{v}$ valence quarks. At the initial input scale $Q_{0}^{2}$, there are no sea quarks or gluon distributions. In other words, we assume that at such low scales, exotic hadrons solely comprise valence quarks. Consequently, the valence sum rule for the naive nonperturbative input is given by:
\begin{equation}
\int_0^1 f_v(x,Q_0^2)dx=n_{v}.
\label{ValenceSum_hadron}
\end{equation}
Since the premise here is the absence of sea quarks and gluons in the initial non-perturbative input, valence quarks account for the total momentum of the exotic hadron. The momentum sum rule for valence quarks at $Q_0^2$ can be expressed as:
\begin{equation}
\int_0^1 x[n_{v}f_v(x,Q_0^2)]dx=1.
\label{MomentumSum_hadron}
\end{equation}

With the constraints in Eqs. (\ref{ValenceSum_hadron}) and (\ref{MomentumSum_hadron}), only one free parameter remains for the nonperturbative input in Eq. (\ref{Parametrization_hadron}). To determine this free parameter, the Maximum Entropy Method (MEM) is employed. Based on the definition of generalized information entropy, the entropy S of valence quark distributions in exotic hadrons is computed as:
\begin{equation}
\begin{aligned}
S=&-\int_0^1 n_{v}\left [ f_v(x,Q_0^2){\rm Ln}(f_v(x,Q_0^2)) \right ] dx.
\end{aligned}
\label{EntropyDefinition_hadron}
\end{equation}
The optimal parameterized initial valence quark distributions for exotic hadrons are derived when the entropy $S$ reaches its maximum value. By utilizing Eqs. (\ref{ValenceSum_hadron}), (\ref{MomentumSum_hadron}), and (\ref{EntropyDefinition_hadron}), we calculate the initial valence quark distributions for hadrons with valence quark numbers $n_{v}$ = 4, 5, and 6.

\begin{figure}[htp]
\begin{center}
\includegraphics[width=0.46\textwidth]{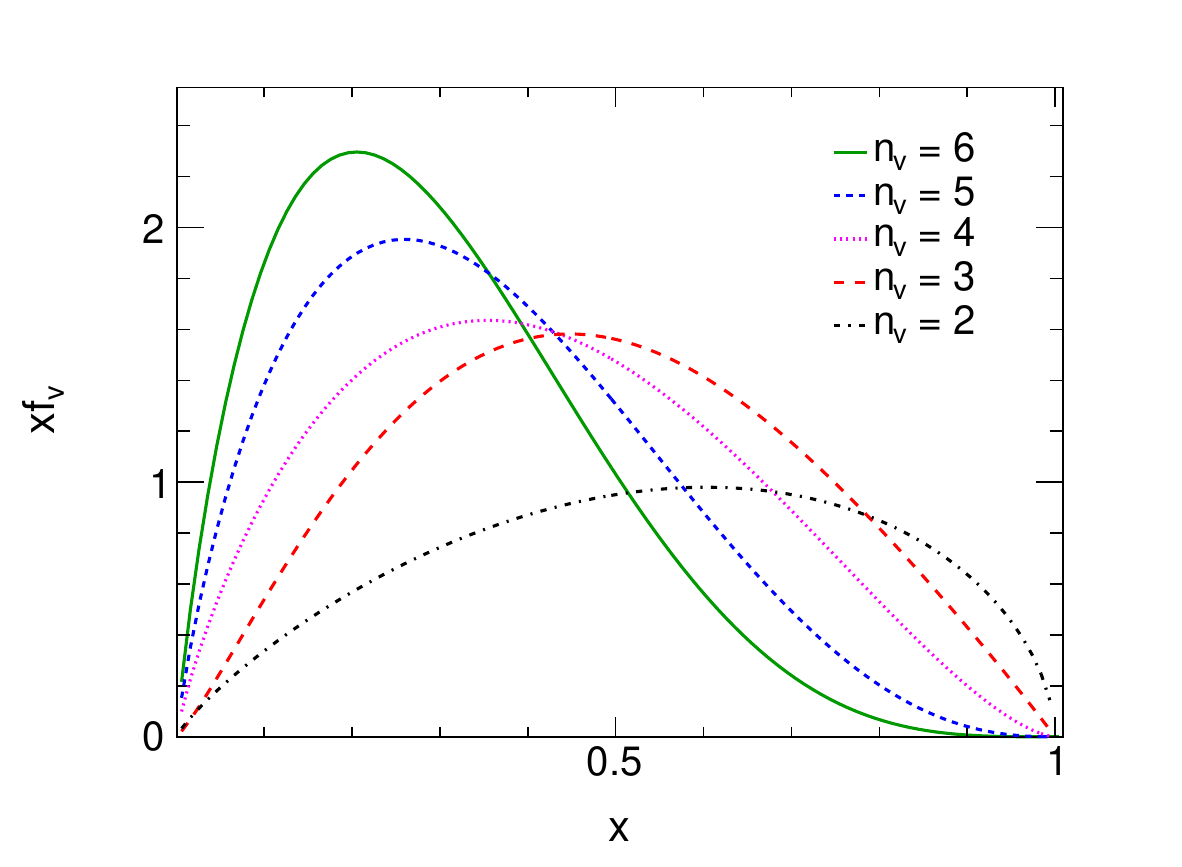}
\caption{~(Color online)
  The valence quark momentum distributions of exotic hadrons in comparison with parametrizations of pion \cite{Han:2018wsw}
  and proton \cite{Wang:2014lua} valence quark distributions previously obtainted by MEM at initial scale $Q_{0}^{2}$.
  Where the n$_{v}$ represents hadrons containing different quark numbers.
}
\label{xfv_x_hadrons}
\end{center}
\end{figure}

Figure \ref{xfv_x_hadrons} displays the valence quark momentum distributions of exotic hadrons, comparing them with the parametrizations of pion \cite{Han:2018wsw} and proton \cite{Wang:2014lua} valence quark distributions previously obtained using MEM at the initial scale $Q_{0}^{2}$. Here, n$_{v}$ represents the number of constituents in the hadrons. From the figure, it is evident that the peak position of the valence quark distribution shifts towards smaller Bjorken scale $x$ as the number of valence quarks in the hadrons increases. The valence quark distributions of the hadrons exhibit peaks at approximately $x=1/n_{v}$, which is in line with the calculations by Kawamura and Kumano \cite{Kawamura:2013wfa}.

\section{A non-perturbative input for the $Z_c(3900)$ exotic hadron}
\label{SecIII}

In our analysis, the valence quark distribution functions at $Q_0^2$ are parametrized to approximate the analytical solution of non-perturbative QCD. The simplest functional form used to approximate the valence quark distribution function is the well-established canonical parametrization $f(x) = A x^B (1-x)^C$ \cite{Pumplin:2002vw}. The observation of the exotic quarkonium state $Z_{c}(3900)$ by the BESIII and Belle Collaborations \cite{BESIII:2013ris,Belle:2013yex,Xiao:2013iha,Wilbring:2013cha} reveals a composition of four valence quarks ($u\bar{d}c\bar{c}$), consisting of three positively charged quarks ($u\bar{d}c$) and one negatively charged quark  ($\bar{c}$). The three positively charged quarks ($u\bar{d}c$) are assumed to share the same spatial region, while the negatively charged anti-charm quark ($\bar{c}$) occupies the entire space, inspired by Ref. \cite{Wang:2014lua}.
Therefore, the simplest parametrization of the naive non-perturbative input is expressed as:
\begin{equation}
\begin{aligned}
&u_v(x,Q_0^2)=\bar{d}_v(x,Q_0^2)=c_v(x,Q_0^2)=A_u x^{B_u}(1-x)^{C_u},\\
&\bar{c}_v(x,Q_0^2)=A_{\bar{c}} x^{B_{\bar{c}}}(1-x)^{C_{\bar{c}}}.
\end{aligned}
\label{Parametrization_zc3900}
\end{equation}

The exotic quarkonium state $Z_{c}(3900)$ consists of $u$, $\bar{d}$, and $c$ valence quarks, which are positively charged, along with a $\bar{c}$ valence quark that is negatively charged. Therefore, the valence sum rule for the naive non-perturbative input can be expressed as:
\begin{equation}
\int_0^1 u_v(x,Q_0^2)dx=3, \quad \int_0^1 \bar{c}_v(x,Q_0^2)dx=1.
\label{ValenceSum}
\end{equation}
Given that there are no sea quarks and gluons in the naive non-perturbative input, the valence quarks carry the total momentum of the $Z_{c}(3900)$. The momentum sum rule for the valence quarks at $Q_0^2$ is satisfied by the following equation:
\begin{equation}
\int_0^1 x[u_v(x,Q_0^2)+\bar{c}_v(x,Q_0^2)]dx=1.
\label{MomentumSum}
\end{equation}

According to the Partons model, the unpolarized structure function of $Z_{c}(3900)$ is written as:
\begin{equation}
2xF_{1}(x)=F_{2}(x)=\sum_{i} e_{i}^{2}xf_{i}(x).
\label{F1}
\end{equation}
This expression is known as the Callan-Gross relational expression \cite{Callan:1969uq}. Here, $i$ represents the flavor index, $e_{i}$ is the electric charge of the quark of flavor $i$ (in units of the electron charge), and $xf_{i}$ denotes the momentum fraction of the quark of flavor $i$.

Color confinement is a characteristic feature of strong interactions in QCD, indicating that all quarks and gluons are confined within a small region of space. Quark confinement is a fundamental aspect of Non-Abelian gauge field theory. According to the Heisenberg uncertainty principle, the momenta of the four valence quarks in the $Z_c(3900)$ are uncertain, implying that the valence quarks have probability density distributions.
The Heisenberg uncertainty principle is expressed as:
\begin{equation}
\sigma_X\sigma_P \ge \frac{\hbar}{2}.
\label{Uncertainty}
\end{equation}
For a quantum harmonic oscillator in the ground state in quantum mechanics, the uncertainty relation is $\sigma_X\sigma_P = \hbar/2$. To constrain the standard deviation of the quark momentum distributions, we assume $\sigma_X\sigma_P = \hbar/2$ for the initial valence quarks in our analysis, rather than $\sigma_X\sigma_P > \hbar/2$.

$\sigma_X$ represents the uncertainty in the radius ($R$) of the $Z_c(3900)$. In this context, we only consider the electromagnetic interaction between the quarks. An approximate estimation involves transforming the spherical $Z_c(3900)$ into a cylindrical shape. One can determine $\sigma_X = (2\pi R^3/3)/(\pi R^2)=2R/3$. The uncertainty $\sigma_X$ for each valence quark is divided by $3^{1/3}$ since there are three positively charged quarks ($u\bar{d}c$) in the spatial region. Therefore, we obtain $\sigma_{X_u}=2R/(3\times 3^{1/3})$ and $\sigma_{X_d}=2R/3$.
The Bjorken variable $x$ represents the momentum fraction of one parton with respect to the $Z_{c}(3900)$. Therefore, we define the standard deviation of $x$ at an extremely low resolution scale $Q_0^2$ as:
\begin{equation}
\sigma_x = \frac{\sigma_P}{M}.
\label{xDeviation}
\end{equation}
Here, $M$ denotes the mass of $Z_{c}(3900)$, which is 3.9 GeV. This work adopts natural units for calculations. Finally, constraints on valence quark distribution functions resulting from quark confinement and the Heisenberg uncertainty principle are provided as follows:
\begin{equation}
\begin{aligned}
&\sqrt{<x_u^2>-<x_u>^2}=\sigma_{x_u},\\
&\sqrt{<x_{\bar{c}}^2>-<x_{\bar{c}}>^2}=\sigma_{x_{\bar{c}}},\\
&<x_u>=\int_0^1 x\frac{u_v(x,Q_0^2)}{3}dx,\\
&<x_{\bar{c}}>=\int_0^1 x\bar{c}_v(x,Q_0^2)dx,\\
&<x_u^2>=\int_0^1 x^2\frac{u_v(x,Q_0^2)}{3}dx,\\
&<x_{\bar{c}}^2>=\int_0^1 x^2\bar{c}_v(x,Q_0^2)dx.\\
\end{aligned}
\label{sigma_xDeviation}
\end{equation}

\section{Determination of $Z_{c}(3900)$ valance quark distributions from maximum entropy method}

\begin{figure}[htp]
\begin{center}
\includegraphics[width=0.46\textwidth]{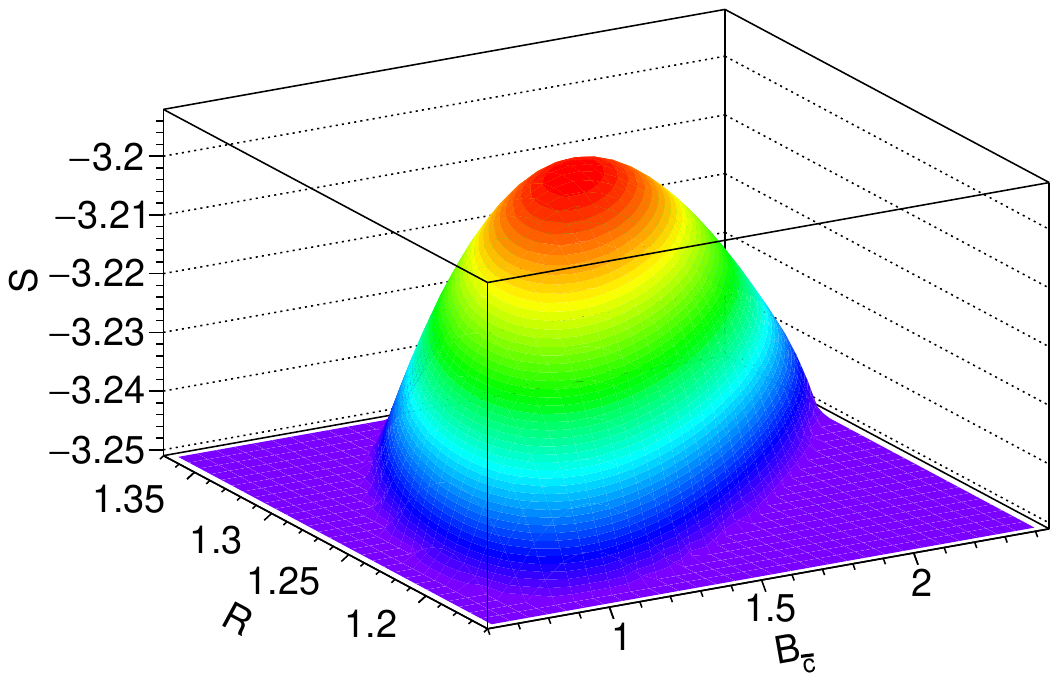}
\caption{~(Color online)
Information entropy $S$ is plotted as a function of the parameters $B_{\bar{c}}$ and $R$.
}
\label{Entropy}
\end{center}
\end{figure}

\begin{figure}[htp]
\begin{center}
\includegraphics[width=0.46\textwidth]{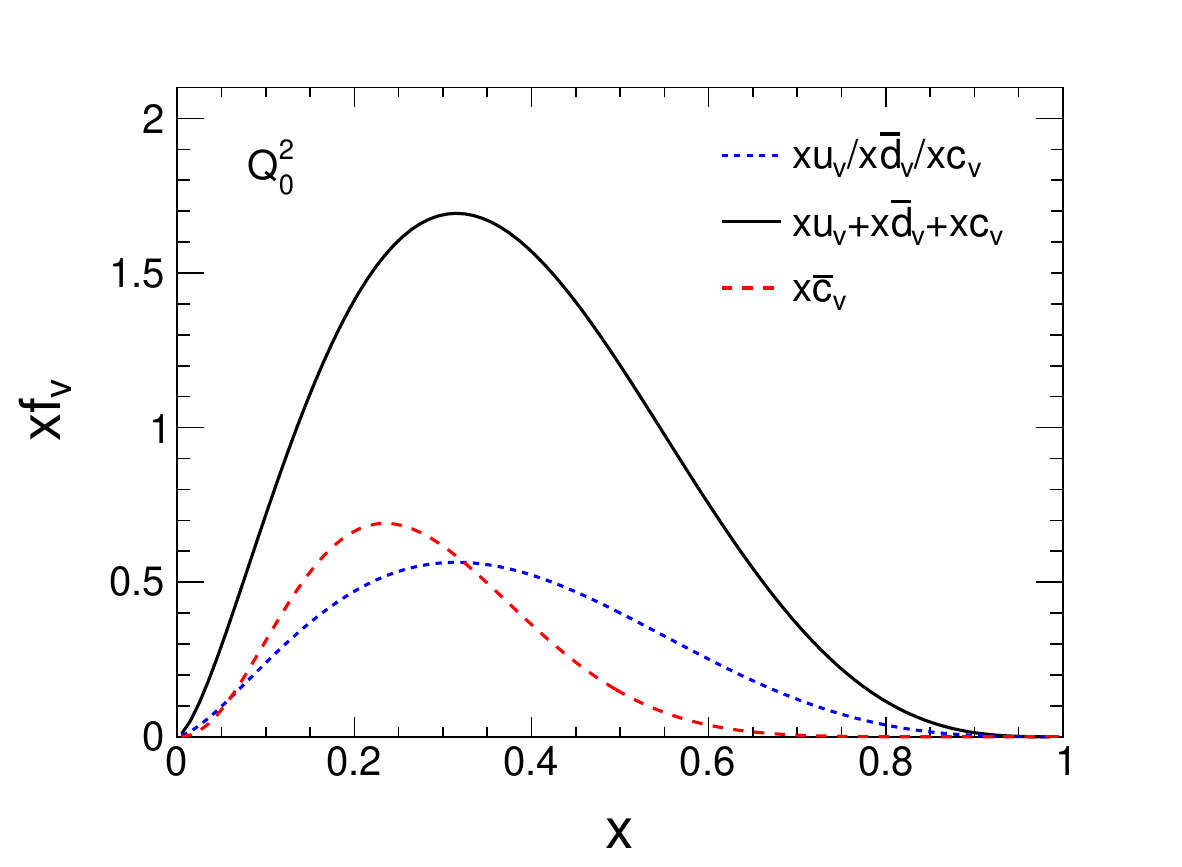}
\caption{~(Color online)
The valence quark momentum distributions of the $Z_{c}(3900)$ at initial scale $Q^{2}_{0}$.
}
\label{valence_zc3900}
\end{center}
\end{figure}

\label{SecIV}
Much of the available evidence suggests that information theory entropy provides a constructive criterion for a given maximum probability distributions under some known constraints, and bring about a type of statistical inference which is called the maximum entropy estimation \cite{Jaynes:1957zza}. It will give us the least biased estimate possible parameter values on the given information. The information theory entropy $S$ \cite{von1995maximum,dembo1991information} is as follows,
\begin{equation}
\begin{aligned}
S=\sum_{i}A_i-m_i-A_{i}ln(\frac{A_i}{m_i}).
\end{aligned}
\label{Entropy_generalized}
\end{equation}
which is relative to the default model $m_{i}$. Here, the default model can be considered as a weak constraint\cite{Asakawa:2000tr}.

By applying MEM, one can find the most reasonable valence quark distributions from the testable information which are the constraints disscussed above.
The generalized information entropy of valence quarks is defined as,
\begin{equation}
\begin{aligned}
  S=&\int_0^1 [3(\frac{u_v(x,Q_0^2)}{3}-Mx(1-x)\\
    &-{\frac{u_v(x,Q_0^2)}{3}}ln(\frac{u_v(x,Q_0^2)}{3Mx(1-x)})) \\
    &+ ({\bar{c}}_v(x,Q_0^2)-Mx(1-x) \\
    &- {\bar{c}}_v(x,Q_0^2)ln(\frac{{\bar{c}}_v(x,Q_0^2))}{Mx(1-x)}) ] dx.
\end{aligned}
\label{Entropy_specific}
\end{equation}
where the $M$ denote mass of $Z_{c}(3900)$. The best estimation of the non-perturbative input parameters in Eq. (\ref{Parametrization_zc3900})
will correspond to the maximum entropy value.
That is to say valence quark distributions are determined by taking the maximum entropy.

We consider the information entropy $S$($B_{\bar{c}}$, $R$) as a function of the variable $B_{\bar{c}}$ and the radius $R$ of $Z_{c}(3900)$. With the constraints
given in Eqs. (\ref{ValenceSum}), (\ref{MomentumSum}), (\ref{Uncertainty}), (\ref{xDeviation}) and (\ref{sigma_xDeviation}),
there are two free parameters left for the parametrized naive non-perturbative input.
Figure.~\ref{Entropy} shows the information entropy $S$ is plotted as a function of the parameters $B_{\bar{c}}$ and $R$.
By selecting the maximum value of entropy, $B_{\bar{c}}$ is optimized to be 1.465 and $R$ is optimized at 1.276 fm.
The corresponding valence quark distribution functions are obtained as follows,
\begin{equation}
\begin{aligned}
&u_v(x,Q_0^2)=36.057x^{0.547}(1-x)^{3.360},\\
&{\bar{c}}_v(x,Q_0^2)=211.309x^{1.465}(1-x)^{8.043}.
\end{aligned}
\label{InitialValence}
\end{equation}
Where the Fig. \ref{valence_zc3900} shows the $Z_{c}(3900)$ valence quarks momentum distributions we obtained.

\begin{figure}[htp]
\begin{center}
\includegraphics[width=0.46\textwidth]{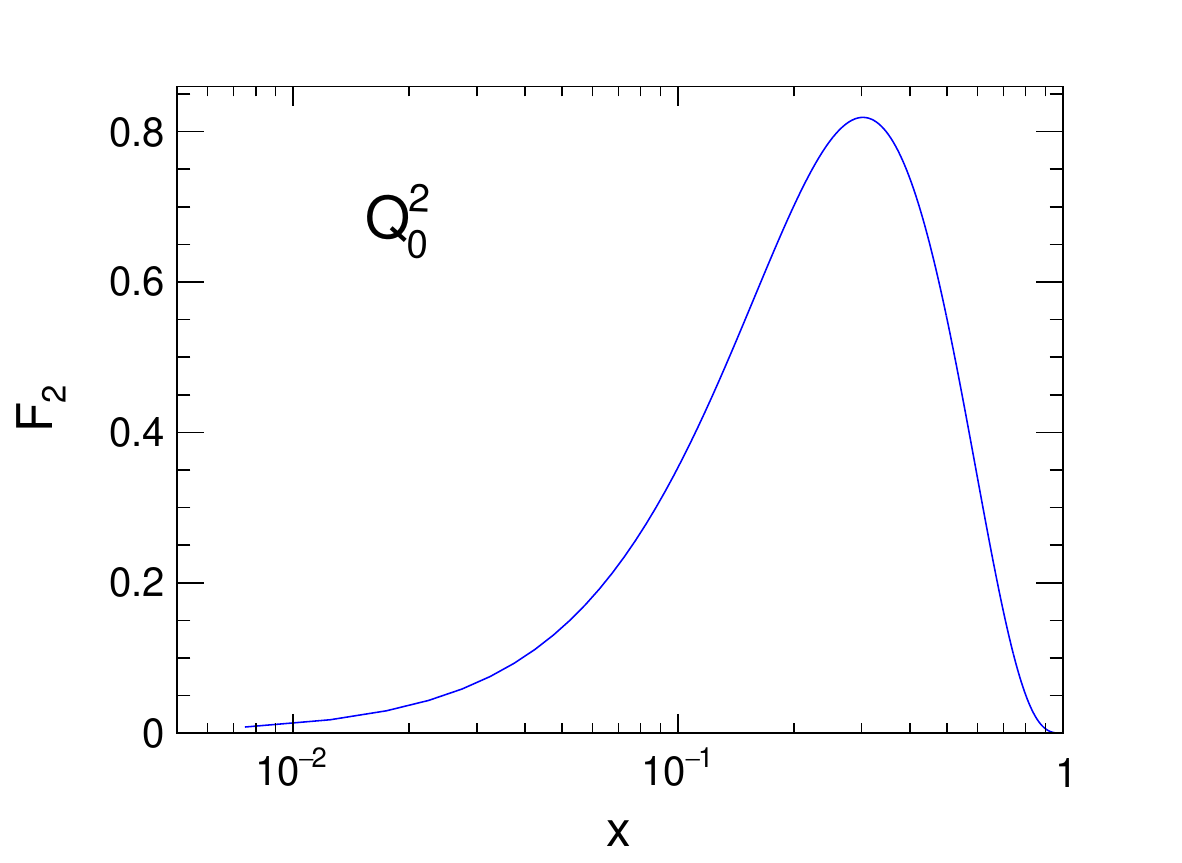}
\caption{~(Color online)
The obtained unpolarized structure function $F_2$ of $Z_{c}(3900)$ as a function of bjorken variable $x$ at initial scale $Q_{0}^{2}$.
}
\label{fig:F2}
\end{center}
\end{figure}

\begin{figure}[htp]
\begin{center}
\includegraphics[width=0.46\textwidth]{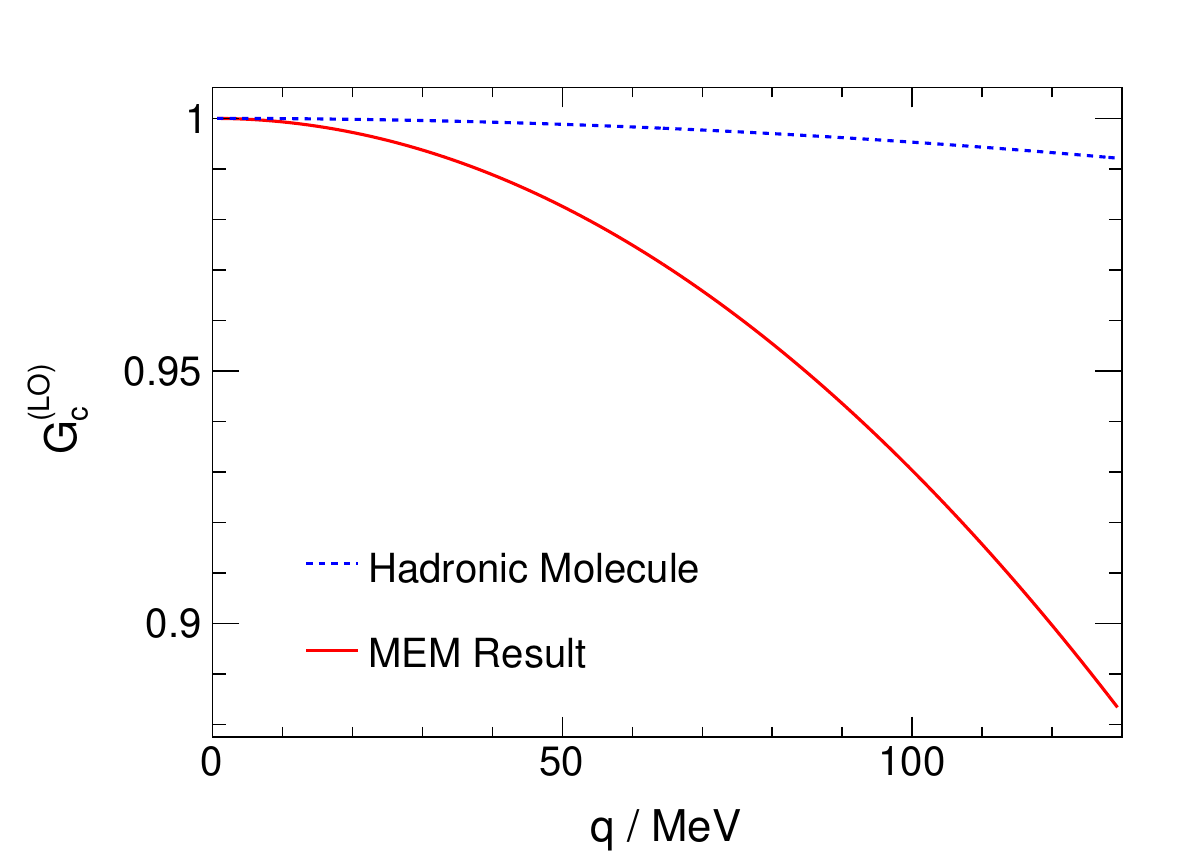}
\caption{~(Color online)
LO charge form factor $G^{LO}_{c}(q)$ of the $Z_{c}(3900)$.
}
\label{Gc_LO}
\end{center}
\end{figure}

The unpolarized structure function $F_2$ of $Z_{c}(3900)$ is essential for understanding the internal structure of exotic hadrons, which is related to quark momentum distributions directly. The corresponding structure function $F_2(x,Q^2_0)$ is written as
\begin{equation}
\begin{aligned}
  F_{2}(x,Q_0^2)=&\frac{4}{9}x[u_v(x,Q_0^2)+c_v(x,Q_0^2)+{\bar{c}}_v(x,Q_0^2)] \\
  &+\frac{1}{9}x{\bar{d}}_v(x,Q_0^2).
\end{aligned}
\label{eq:F2}
\end{equation}
The structure function of $Z_{c}(3900)$ can be calculated through the obtained valence quark distribution function from the MEM. The obtained $F_2$ of $Z_{c}(3900)$ as a function of bjorken variable $x$ at $Q_{0}^{2}$ is shown in Fig. \ref{fig:F2}.

The radius of the $Z_{c}(3900)$ is equal to 1.276 fm, so that the charge form factor of $Z_{c}(3900)$ can be obtained. The leading order (LO) charge form factor $G^{LO}_{c}(q)$ is written as
\begin{equation}
G_{c}(q)=1-<r_c^{2}>q^{2}/6.
\label{Gc}
\end{equation}
where the $<r_c^2>$ is the value of the squared charge radius of $Z_{c}(3900)$. The charge form factor of the $Z_{c}(3900)$ at LO by maximum entropy (solid line) is shown in Figure. 4. Our model shows that the $Z_{c}(3900)$ is a hadronic state. The dashed line is result of hadronic molecule state \cite{Wilbring:2013cha}.
We hope that the charge form factor $G^{LO}_{c}(q)$ obtained by MEM can be verified by future high energy exclusive processes that include generalized parton
distributions and generalized distribution amplitudes.

\section{GDA and exotic hadrons}
When considering the expression of exotic hadrons, the possibilities of compact multiquark states and molecular states exist. Due to the potential to achieve similar masses and decay widths in ordinary hadron models, distinguishing between the various internal configurations becomes challenging. Nevertheless, Kawamura and Kumano \cite{Kawamura:2013wfa} suggested that the internal configurations of exotic hadron candidates could be elucidated through exclusive processes in the high-energy region. Perhaps we can establish a connection between the form factor related to the multi-quark state PDFs mentioned in the previous section and experiments, providing an indirect but phenomenologically relevant approach. In general, it may be possible to link the GDA with the form factor starting from the high-energy scattering cross-section of exclusive processes, thereby testing the multi-quark state distribution estimated using MEM.

\begin{figure}[htbp]
    \centering
    \includegraphics[width=0.35\textwidth]{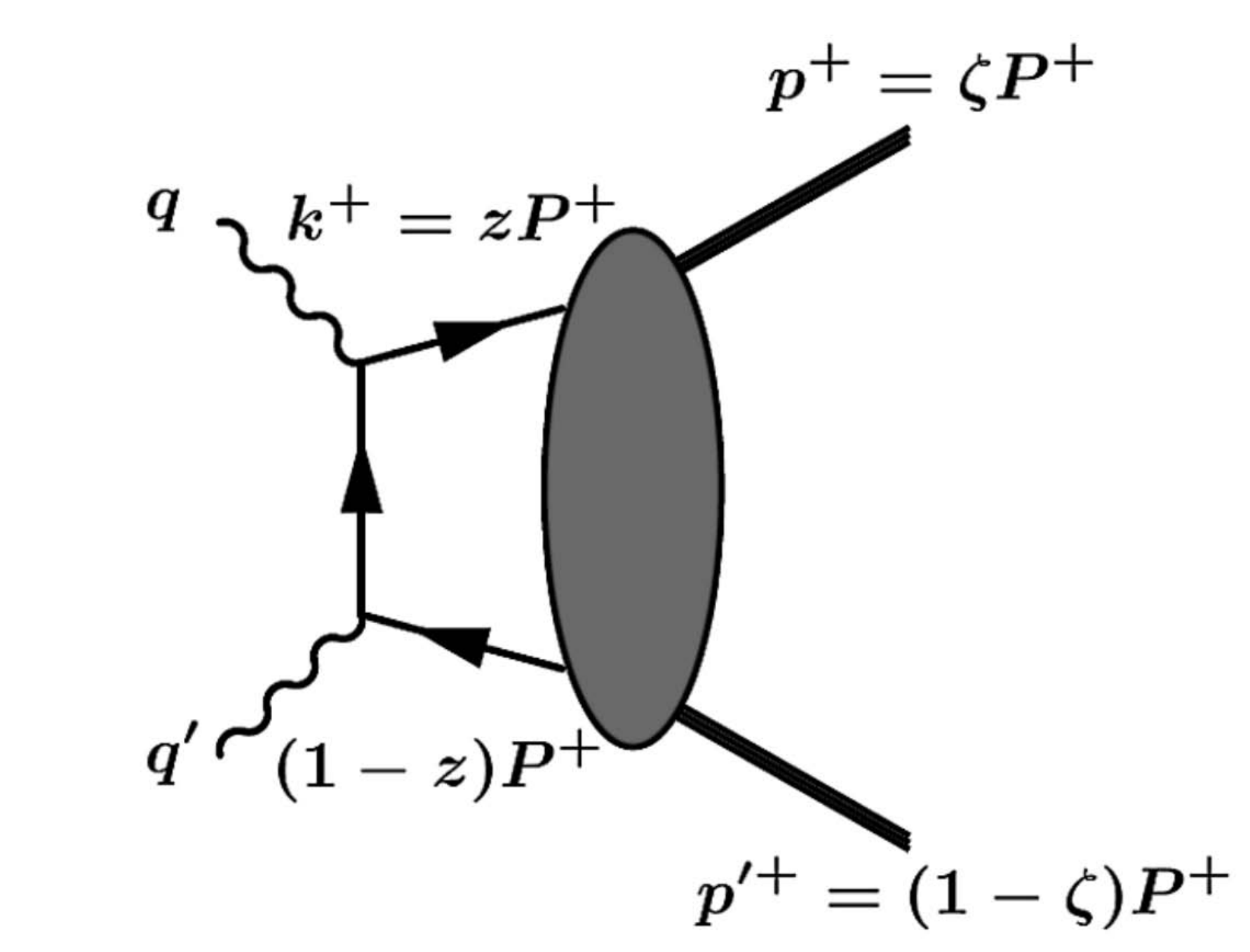}
    \caption{$\gamma^*\gamma\rightarrow h\bar{h}$ process.}
    \label{fig:rrhGDA}
\end{figure}

Exclusive processes is related to generalized parton distributions (GPD) or GDA, which can determine the internal structure of hadron. As shown in Fig.\,\ref{fig:rrhGDA} , in the kinematical region of $Q^2\gg W^2$, the process $\gamma^* \gamma \rightarrow h\bar{h}$ can be factorized into two parts. One is a hard process which is described by $\gamma^* \gamma \rightarrow q\bar{q}$, the other is a soft process which is described by GDA for the production of two hadrons from $q\bar{q}$.

GDA is defined as 
\begin{equation}
\label{eq:general_definition}
\begin{aligned}\Phi_q^{h\bar{h}}(z,\zeta,W^2)&=\int\frac{dx^-}{2\pi}e^{-izP^+x^-}\\&\times\langle h(p)\bar{h}(p^{\prime})|\bar{q}(x^-)\gamma^+q(0)|0\rangle_{x^+=0,\vec{x}_\perp=0},\end{aligned}
\end{equation}
where $z$ and $\zeta$ are given by 
\begin{equation}z=\frac{k\cdot q^{\prime}}{P\cdot q^{\prime}}=\frac{k^+}{P^+},\quad \zeta=\frac{p\cdot q^{\prime}}{P\cdot q^{\prime}}=\frac{p^+}{P^+}
.\end{equation}
The relationship between GDA and GPD is
\begin{equation}
\label{eq:GDA_GPD}
\Phi_q^{h\bar h}(z,\zeta,W^2)\leftrightarrow H_q^h\Bigg(x=\frac{1-2z}{1-2\zeta},\xi=\frac{1}{1-2\zeta},t=W^2\Bigg).
\end{equation}
 In the forward limit, $H_q(x,0,0)=q(x)$, where $q(x)$ is parton distribution function. The difference between the exotic hadron and the compact quark state for $q(x)$ should exist.

\begin{figure}[htbp]
    \centering
    \includegraphics[width=0.46\textwidth]{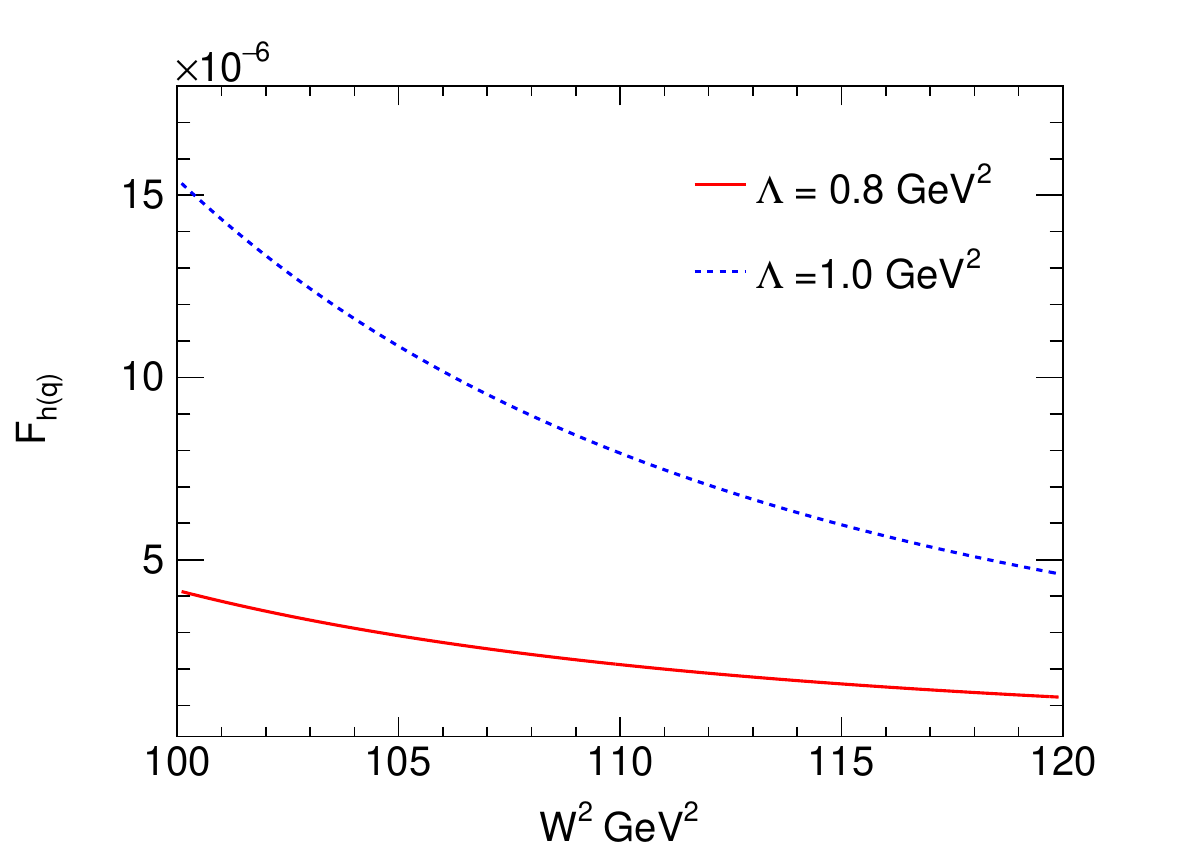}
    \caption{The influence of the $\Lambda$ on the form-factor as a function of the invariant mass squared $W^2$ when $n=4$ and $m_{h}=3.9\ \mathrm{GeV}^2$.}
    \label{fig:diff_cross_section.pdf}
\end{figure}

For the production of the two-meson final state, a simple form of quark GDA with isospin $I=0$ on the two meson final state is 
\begin{equation}
\begin{aligned}\Phi_q^{h\bar{h}(I=0)}(z,\zeta,W^2)&=N_{h(q)}z^\alpha(1-z)^\beta(2z-1)\\&\times\zeta(1-\zeta)F_{h(q)}(W^2),
\end{aligned}
\end{equation}
where $F_{h(q)}(W^2)$ is form factor and $N_{h(q)}$ is defined as
\begin{equation}
\label{eq:Nhq}
\begin{aligned}
&N_{h(q)}=-\frac{2M_{2(q)}^h}{B(\alpha+1,\beta+1)}\\&\times\frac{(\alpha+\beta+2)(\alpha+\beta+3)}{(\alpha+1)(\alpha+2)+(\beta+1)(\beta+2)-2(\alpha+1)(\beta+1)},
\end{aligned}
\end{equation}
where $\alpha$ and $\beta$ are constants. In \cite{Kawamura:2013wfa}, $\alpha=\beta=$ 1, 2 and 3 is taken. The form factor $F_{h(q)}(W^2)$ is 
\begin{equation}
\label{eq:F_hq}
F_{h(q)}(W^2)=\frac{1}{[1+(W^2-4m_h^2)/\Lambda^2]^{n-1}},
\end{equation}
$m_h$ is the mass of the hadron. $\Lambda$ is the QCD cutoff parameter, which signifies the hadron size, $n$ is the number of constituent quarks. $\Lambda$ is an important variable to distinguish molecular state and compact multiquark state, beacuse the size of them are different. As shown in Fig.\,\ref{fig:diff_cross_section.pdf}, the parameter $\Lambda$ plays an important role on the form factor.  
Kawamura and Kumano \cite{Kawamura:2013wfa} also presented $\Lambda$ is important on $\frac{d\sigma_{e\gamma\rightarrow eh\overline{h}}}{dQ^2dW^2}$.  

Although a direct measurement method for the longitudinal PDF of multi-quark states has not yet emerged, the measurement scheme for GDA has been widely discussed. The MEM proposed by us can be used as an input to estimate the internal structure of multi-quark states, such as GPD and GDA, and can also serve as a reference for future experimental measurements. In addition, the increase in the number of quarks in hadrons leading to a shift of the peak of the low-energy longitudinal PDF towards smaller momentum fractions is also reasonable. The distribution of hadron momentum depends on the number of quarks, and the larger the number of quarks, the softer the quark distribution.

\section{summary}
\label{SecV}
In this work, using the valence sum rules and the momentum sum rule and Eq. ( \ref{EntropyDefinition_hadron}), we calculate the initial valence quark distributions for
hadrons with the valence quark number $n_{v}$ = 4, 5 and 6 from the MEM.
From the calculation results, we can find that the peak position of the valence quark distribution of hadrons containing more valence quarks moves toward smaller $x$.
The valence quark momentum distributions of hadrons have peaks at about $x=1/n_{v}$, which is consistent with the results of Kawamura and Kumano \cite{Kawamura:2013wfa}.

Furthermore, this is a very meaningful attempt of determining the $Z_{c}(3900)$ radius and parton distribution functions from the MEM.
In this work, we take the information entropy S($B_{\bar{c}}$, R) as a function of the variable $B_{\bar{c}}$ and the radius $R$ of $Z_{c}(3900)$.
Under the constraint of the equation above, there are two free parameters left for the parametrized naive nonperturbative input.
After determining the unknown free parameters, the valence quark distribution function are given accordingly.
One can further get structure function at $Q^2_0$. The radius ($R$) of the $Z_{c}(3900)$ is equal to 1.276 fm,
so the leading order (LO) charge form factor $G^{LO}_{c}(q)$ of $Z_{c}(3900)$ can be obtained.
By comparison, the form factor $G^{LO}_{c}(q)$ of $Z_{c}(3900)$ we calculated by MEM has an apparent difference from hadron molecular state.
We suggest that the charge form factor $G^{LO}_{c}(q)$ obtained by MEM can be verified by future high energy exclusive processes that include generalized parton
distributions and generalized distribution amplitudes.

Future experiments at high-energy facilities, such as the J-PARC and the upcoming Electron Ion Collider (EIC) in the United States \cite{Aschenauer:2014twa, Aguilar:2019teb, Arrington:2021biu}, will provide a valuable opportunity to test the obtained valence quark distributions of exotic hadrons. The EIC, currently under construction, and the proposed Electron-Ion Collider in China (EicC) \cite{Anderle:2021wcy, Chen:2018wyz, Chen:2020ijn} will also offer a unique platform to validate our findings.

\begin{acknowledgments}
This work is supported by the National Natural Science Foundation of China under the Grant No. 12305127,
the International Partnership Program of the Chinese Academy of Sciences under the Grant No. 016GJHZ2022054FN,
the Strategic Priority Research Program of Chinese Academy of Sciences under Grant No. XDB34030301 and
the Guangdong Major Project of Basic and Applied Basic Research No. 2020B0301030008.
\end{acknowledgments}

\bibliographystyle{apsrev4-1}
\bibliography{refs.bib}

\end{document}